\documentclass[english,aps,prl,notitlepage,twocolumn]{revtex4-1}
\usepackage[T1]{fontenc}
\usepackage{color}
\usepackage{amsmath}
\usepackage{amssymb}
\usepackage{graphicx}

\makeatletter

\usepackage{amsmath}
\usepackage{endnotes}
\newcommand\ket[1]{\left| #1\right\rangle}
\newcommand\bra[1]{\left\langle #1\right|}
\newcommand\braket[2]{\left\langle #1| #2\right\rangle }

\makeatother

\usepackage{babel}
\begin{document}
\global\long\def\ket#1{\left| #1\right\rangle }

\global\long\def\bra#1{\left\langle #1\right|}

\global\long\def\braket#1#2{\left\langle #1| #2\right\rangle }

\global\long\def\mk{\bm{k}}

\global\long\def\mr{\bm{r}}

\title{Flat-band localization in Creutz superradiance lattices}

\author{Yanyan He,$^{1,*}$ Ruosong Mao,$^{1,*}$ Han Cai,$^{1,\dagger}$ Jun-Xiang Zhang,$^{1,\dagger\dagger}$ Yongqiang Li,$^{2}$ Luqi Yuan,$^{3}$ Shi-Yao Zhu,$^1$ and Da-Wei Wang$^{1,4}$}

\affiliation{$^1$Interdisciplinary Center for Quantum Information, State Key Laboratory of Modern Optical Instrumentation, and Zhejiang Province Key Laboratory of Quantum Technology and Device, Department of Physics, Zhejiang University, Hangzhou 310027, Zhejiang Province, China;\\
$^2$Department of Physics, National University of Defense Technology, Changsha 410073, Hunan Province, China;\\
$^3$State Key Laboratory of Advanced Optical Communication Systems and Networks, School of Physics and Astronomy, Shanghai Jiao Tong University, Shanghai 200240, China;\\
$^4$CAS Center for Excellence in Topological Quantum Computation, University of Chinese Academy of Sciences, Beijing 100190, China
}

\date{\today}
\begin{abstract}
Flat bands play an important role in diffraction-free photonics and
attract fundamental interest in many-body physics. Here we report
the engineering of flat-band localization of collective excited states
of atoms in Creutz superradiance lattices with tunable synthetic gauge
fields. Magnitudes and phases of the lattice hopping coefficients
can be independently tuned to control the state components of the
flat band and the Aharonov-Bohm phases. We can selectively excite
the flat band and control the flat-band localization with the synthetic
gauge field. Our study provides a room-temperature platform for flat
bands of atoms and holds promising applications in exploring correlated
topological materials.
\end{abstract}
\maketitle
Flat bands are characterized by the zero bandwidth over the whole
Brillouin zone. Owing to the destructive interference between the
hopping pathways \citep{sutherland1986localization,bergman2008band},
the group velocity of excitations vanishes, and hence the diffusion
in flat bands is inhibited. The resulting compact localized eigenstates
(CLSs) \citep{flach2014detangling,leykam2018artificial} have been
experimentally realized in photonics \cite{vicencio2015,mukherjee2015a,mukherjee2015b,mukherjee2017,mukherjee2018,kremer2020,xia2016,xia2018,ma2020}
and polariton-exciton condensates \cite{baboux2016,harder2020}. Immune
to environmental noises, localized states in flat bands are promising
candidates for realizing quantum networks \cite{rontgen2019} and
diffraction-free photonics \cite{vicencio2013,rojas2017,yu2020}.
Flat bands are also of fundamental interest in many-body physics because
of their high degeneracy. The density of states is divergent such
that even weak interactions lead to strong correlations and exotic
topological phases \cite{bergholtz2013,junemann2017exploring,spanton2018,chen2020}.

\begin{figure}
\includegraphics[width=0.9\columnwidth]{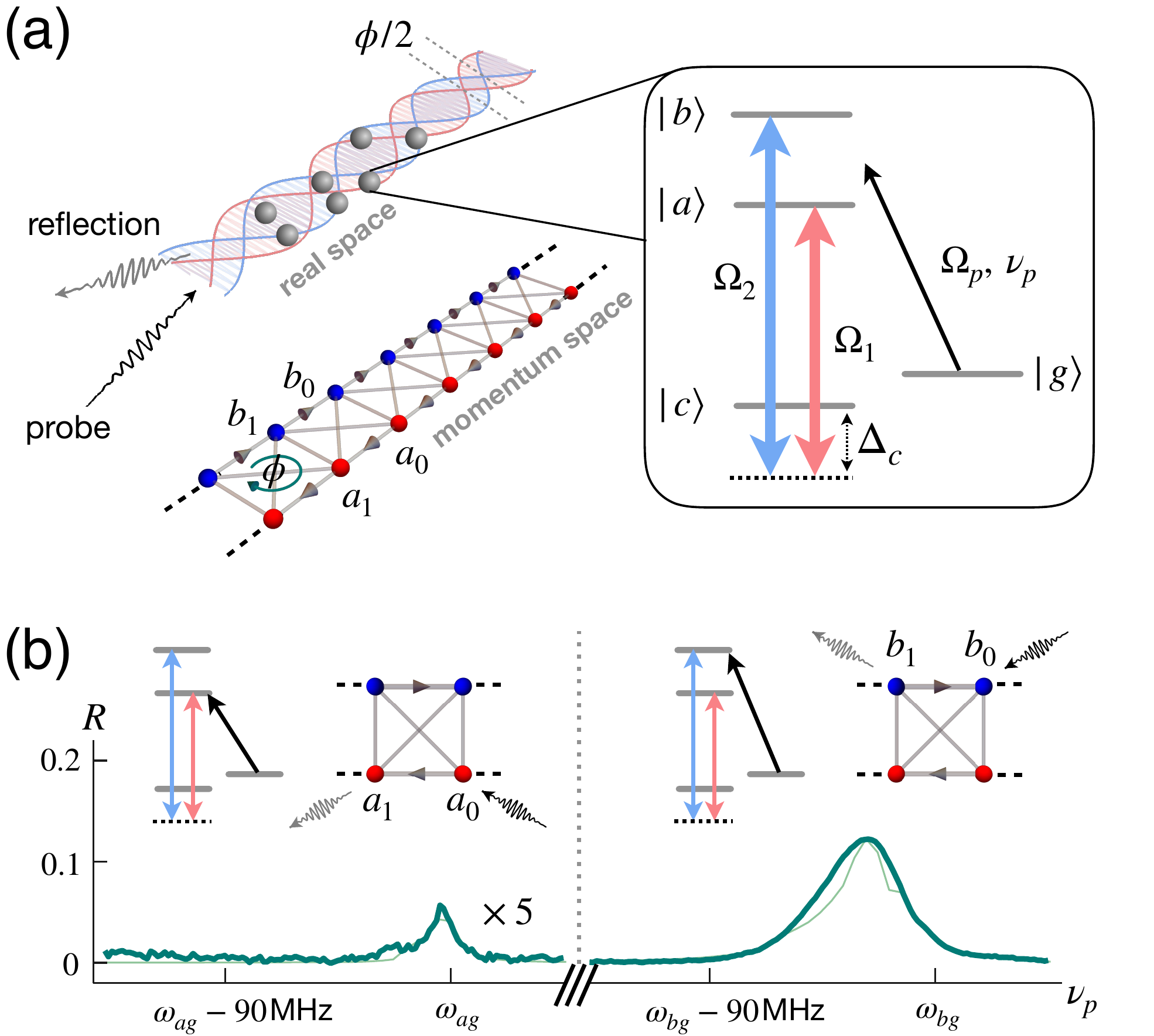}

\caption{(a) Schematic configuration of the experiment. The amplitude envelopes
of the two standing waves have a $\phi/2$ phase difference. In the
Creutz ladder, the arrows indicate the phase $\phi/2$ attached to
the transitions. Inset is the configuration of atomic levels and laser
fields. (b) Typical reflection spectrum. The insets shows the configurations
of the fields and the lattice responses when $\nu_{p}$ is near resonant
with either atomic transitions. The AB flux $\phi=\pi$. The power
of the probe field is $24\text{ \ensuremath{\mu}W}$. The powers of
each plane wave component of the two standing waves are $P_{1}=29\text{ mW}$
and $P_{2}=215\text{ mW}$ with effective Rabi frequencies $\Omega_{1}=15\text{ MHz}$
and $\Omega_{2}=68\text{ MHz}$. The thick dark (thin light) lines
are the experimental data (numerical simulation).}
\label{fig1}
\end{figure}

Many-body interactions can be engineered to realize correlated topological
phases in atoms \cite{bloch2008,li2020}. However, in previous realizations
of the flat bands in optical lattices, the underlying lattices \cite{jo2011,taie2015}
are gapless and topologically trivial. A feasible model that integrates
both band flatness and topology is the two-leg ladder in a uniform
magnetic field with cross-linked couplings, i.e., the Creutz lattice
\cite{Creutz1999End,hugel2014} (see Fig.$\ $\ref{fig1}(a)). Despite
theoretical proposals in photonic waveguides \cite{mukherjee2018}
and ultracold atoms \citep{junemann2017exploring}, flat bands in
the Creutz lattice have never been experimentally realized \cite{kang2018,alaeian2019,kang2020}.

Here we report the synthesis of a Creutz ladder with tunable tight-binding
parameters in the form of a momentum-space lattice, i.e., the superradiance
lattice \citep{WangPRL2015,Chen2018}, in room-temperature cesium
atoms. The bipartite ladder consists of timed Dicke states with different
momenta \citep{ScullyPRL2006}. We find that the corresponding energy
band structure exhibits a flat band and a dispersive band, which are
distinguished by localized and delocalized excitations, respectively.
In the experiment, we excite one site in the ladder with a weak probe
field and measure the optical response of the adjacent site. The hopping
strengths and the Aharonov-Bohm (AB) phases in the lattice are carefully
tuned, which enables us to excite a particular band and control the
flat-band localization. We observe that the optical response is significantly
suppressed when the flat band is selectively excited. By controlling
the AB phases, we reveal the relation between the flat-band localization
and gauge fields \cite{khomeriki2016}. Our work demonstrates a versatile
platform for flat bands of atoms with multiple tunable parameters,
which holds promising applications in exploring correlated topological
phases.

We first introduce the experimental scheme implemented in the hyperfine
levels of the $^{133}$Cs D1 line in a bichromatic standing-wave-coupled
configuration, as shown in Fig.$\ $\ref{fig1}(a). Two standing waves
couple 
two excited states $\ket a\equiv\ket{6^{2}P_{1/2},F=3}$ and $\ket b\equiv\ket{6^{2}P_{1/2},F=4}$
to the same metastable state $\ket c\equiv\ket{6^{2}S_{1/2},F=3}$.
The frequency of the $j$th standing-wave coupling field $\nu_{j}$
fulfills the two-photon resonance condition $\Delta_{c}=\nu_{1}-\omega_{ac}=\nu_{2}-\omega_{bc}$,
where $\omega_{ij}$ being the atomic transition frequency between
$\ket i$ and $\ket j$. The envelopes of the Rabi frequency amplitude
of the two standing waves are $2\Omega_{1}\cos(k_{c}x-\phi/4)$ and
$2\Omega_{2}\cos(k_{c}x+\phi/4)$, where $k_{c}$ is the $x$ component
of the wave vectors and $\phi/2$ is the phase difference between
the envelopes. The wave-vector difference between the two standing
waves is negligible in the length of the atomic vapor cell. We use
a weak travelling field with the wave vector $k_{p}$ to probe the
standing-wave-coupled atomic vapor and measure the backward reflection.
The frequency of the probe field $\nu_{p}$ is scanned to couple the
ground state $\ket g\equiv\ket{6^{2}S_{1/2},F=4}$ to either the state
$\ket a$ or $\ket b$. Featured signals can be observed when the
probe field is near resonant with each atomic transition. A typical
spectrum is shown in Fig.$\ $\ref{fig1}(b). 

In order to show that our experiment constructs a Creutz ladder and
reveal the connection between the reflection signal and the excitation
transport in the ladder, we write the Hamiltonian $H=H_{s}+H_{p}$
in momentum space \cite{supp}, with $H_{s}$ and $H_{p}$ being the
parts of the Hamiltonian corresponding to the couplings of standing
waves and the probe field, respectively. Here, we set $\hbar=1$ and
$H_{s}$ reads

\begin{align}
H_{s} & =\sum_{n}[2t_{1}a_{n}^{\dagger}a_{n}+2t_{2}b_{n}^{\dagger}b_{n}\nonumber \\
 & +(2t_{3}\cos\frac{\phi}{2}a_{n}^{\dagger}b_{n}+t_{3}a_{n}^{\dagger}b_{n+1}+t_{3}b_{n}^{\dagger}a_{n+1}\nonumber \\
 & +t_{1}e^{-i\phi/2}a_{n}^{\dagger}a_{n+1}+t_{2}e^{i\phi/2}b_{n}^{\dagger}b_{n+1}+h.c.)],\label{eq:lattice}
\end{align}
which gives a tight-binding superradiance lattice composed of the
collective atomic excitation operators $d_{j}^{\dagger}=\sqrt{1/N}\sum_{m}\ket{d_{m}}\bra{g_{m}}\exp[i(k_{p}-2jk_{c})x_{m}]$
($d=a,b$) \citep{WangPRL2015}, where $m$ labels the $m$th atom
at the position $x_{m}$, $j$ is an integer, and $N$ is the total
number of atoms. $t_{1}=-\Omega_{1}^{2}/\Delta_{c}$ and $t_{2}=-\Omega_{2}^{2}/\Delta_{c}$
are the hopping amplitudes along $a$-leg and $b$-leg, respectively.
Here, we can adiabatically eliminate the state $\ket c$, since $\Delta_{c}$
is much larger than all relevant Rabi frequencies $(\Delta_{c}\gg\Omega_{j})$.
The two hoppings acquire a phase $\phi/2$ in opposite directions.
The loop transition along a plaquette accumulates an AB phase $\phi$,
such that the lattice is effectively in a uniform magnetic field.
$t_{3}=-\Omega_{1}\Omega_{2}/\Delta_{c}$ and $2t_{3}\cos\phi/2$
are the hopping strengths along the diagonals and the rungs of each
plaquette in the ladder. The on-site energies of the $a$-leg and
$b$-leg sites are $2t_{1}$ and 2$t_{2}$, respectively.

The probe field coupling Hamiltonian is $H_{p}=\sqrt{N}\Omega_{p}e^{-i\Delta_{p}^{\prime}t}a_{0}^{\dagger}+\sqrt{N}\Omega_{p}^{\prime}e^{-i\Delta_{p}t}b_{0}^{\dagger}+h.c.$,
where $\Omega_{p}$ ($\Omega_{p}^{\prime}$) and $\Delta_{p}=\nu_{p}-\omega_{bg}$
($\Delta_{p}^{\prime}=\nu_{p}-\omega_{ag}$) are the Rabi frequency
and the frequency detuning of the coupling between the probe field
and the atomic transition between $\ket a$ ($\ket b$) and $\ket g$.
Hence, $H_{p}$ shows that the excitation is prepared by the probe
field to the site $a_{0}$ or $b_{0}$ in the ladder. When we probe
the site $a_{0}$ ($b_{0}$), the phase-matching condition $k_{p}-2k_{c}\approx-k_{p}$
is only satisfied for the excitation on the site $a_{1}$ ($b_{1}$),
which results in a superradiant backward emission collected by a photodetector.
The spectrum in the left (right) of Fig.$\ $\ref{fig1}(b) characterizes
the excitation transport from $a_{0}$ to $a_{1}$ ($b_{0}$ to $b_{1}$)
in the ladder of Eq.$\ $(\ref{eq:lattice}). In the experiment, the
probe field is weak $(\Omega_{p}\ll t_{i})$ such that only a small
fraction of the atoms are excited. In this condition, $a_{j}$, $b_{j}$
are approximately bosonic annihilation operators \citep{WangPRL2015}. 

\begin{figure}
\includegraphics[width=0.7\columnwidth]{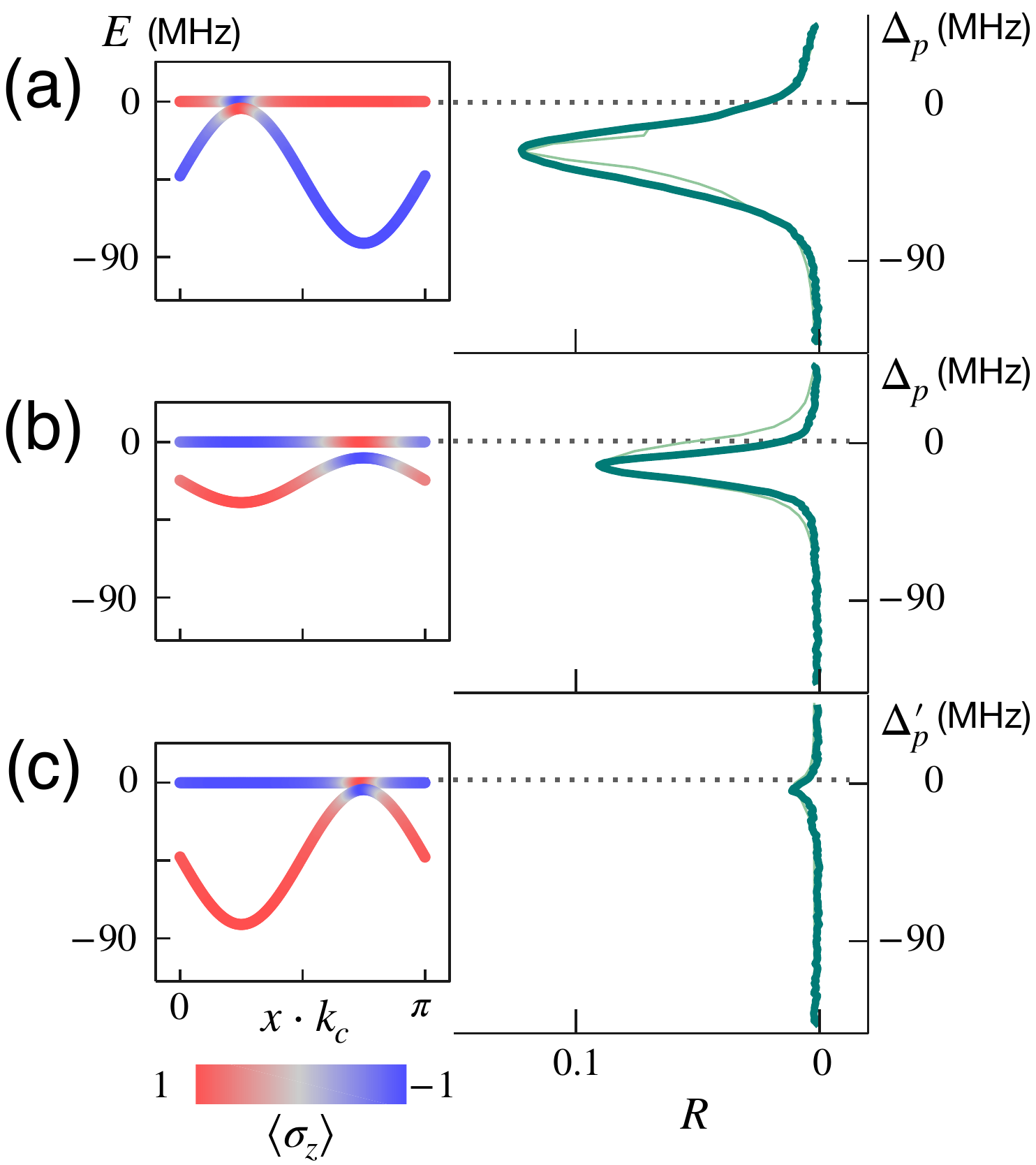}\caption{Band structures (left) and the corresponding reflection spectra (right)
with (a) $\eta=20.7$, (b) $\eta=4.0$, and (c) $\eta=1/20.7$. The
reflection spectrum is mainly contributed by the dispersive (flat)
band for $\eta\gg1$ ($\eta\ll1$). The data in (c) is obtained by
measuring the $a$-leg response owing to the symmetry of the Hamiltonian
(see the text). Dotted gray lines mark the energy of the flat band.}
\label{fig2}
\end{figure}

The Creutz ladder in Eq.$\ $(\ref{eq:lattice}) constructed in momentum
space is tunable in the experiment. We diagonalize $H_{s}$ in real
space \cite{supp} and the band structures are shown in Fig.$\ $\ref{fig2}.
We define $\eta\equiv t_{2}/t_{1}$ as the relative hopping strength
along the two legs. One can see that all three band structures with
different $\eta$ are composed of a flat band and a dispersive band.
In general, the dynamics of the excitation is governed by both bands
and cannot be distinguished. Probing only one band by controlling
the excitation energy \cite{jacqmin2014,drost2017,slot2017} is inapplicable
since the band gap closes when $\phi$ approaches zero \cite{supp}. 

Nevertheless, the tunability of the \textcolor{black}{hopping strengths
enables us to determine which band to excite by controlling the state
component of the bands. In the experiment, we tune $\eta$, which
is proportional to} $P_{2}/P_{1}$, where $P_{1}$ and $P_{2}$ are
the powers of the two standing waves. 
 An interesting correlation can be noticed between the parameter $\eta$
and the band components, where the color represents the polarization
$\left\langle \sigma_{z}\right\rangle \equiv\left\langle \ket a\bra a-\ket b\bra b\right\rangle $
of the eigenstates. In particular, $\left\langle \sigma_{z}\right\rangle =+1$
or $-1$ means the band fully locates on the $a$- or $b$-leg. In
Fig.$\ $\ref{fig2}(a), one can see that $\langle\sigma_{z}\rangle\approx-1$
for almost the whole dispersive band, meaning that the dispersive
band supports a large excitation component on the $b$-leg for $\eta\gg1$
(we take $\eta=20.7$ according to the experimental parameters). Therefore,
the $b_{0}\rightarrow b_{1}$ transport dynamics is governed by the
dispersive band. On the other hand, for $\eta\ll1$ ($\eta=1/20.7$
as shown in \ref{fig2}(c)), $\langle\sigma_{z}\rangle\approx-1$
for almost the entire flat band, so the $b_{0}\rightarrow b_{1}$
transport dynamics is governed by the flat band. 

This band selection is manifested in the bandwidth, the central frequency,
and the magnitude of the reflection spectrum in Fig.$\ $\ref{fig2}.
In the experiment, we change $\eta$ and keep $t_{3}\propto\sqrt{P_{1}P_{2}}$
a constant. In Fig.$\ $\ref{fig2}(a) for $\eta\gg1$, the reflection
spectrum of the dispersive band has a larger bandwidth and a lower
central frequency. As a comparison, in Fig.$\ $\ref{fig2}(c) for
$\eta\ll1$, the reflection spectrum due to the flat band has a much
narrower bandwidth and the peak locates near the predicted frequency
of the flat band. The localization in the flat band is demonstrated
by the decrease of the reflection peak when we decrease $\eta$, during
which the reflection is more contributed by the flat band. 

As a side note, in obtaining Fig.$\ $\ref{fig2}(c), we use the symmetry
that lattice Hamiltonian $H_{s}$ is invariant when we exchange the
sublattices $a$ and $b$, inverse $\eta$, and flip the flux $\phi$
\cite{supp}. In the experiment, the $b_{0}\rightarrow b_{1}$ transport
dynamics with $\phi$ and $\eta>1$ is characterized by the reflection
spectrum near resonant with level $\ket b$ (labelled with $R_{\eta}(\phi)$),
while the one with $-\phi$ and $(1/\eta)<1$ is effectively obtained
from the $\ket a$-side reflection spectrum (labelled with $R_{1/\eta}(-\phi)$). 

\begin{figure}
\includegraphics[width=0.8\columnwidth]{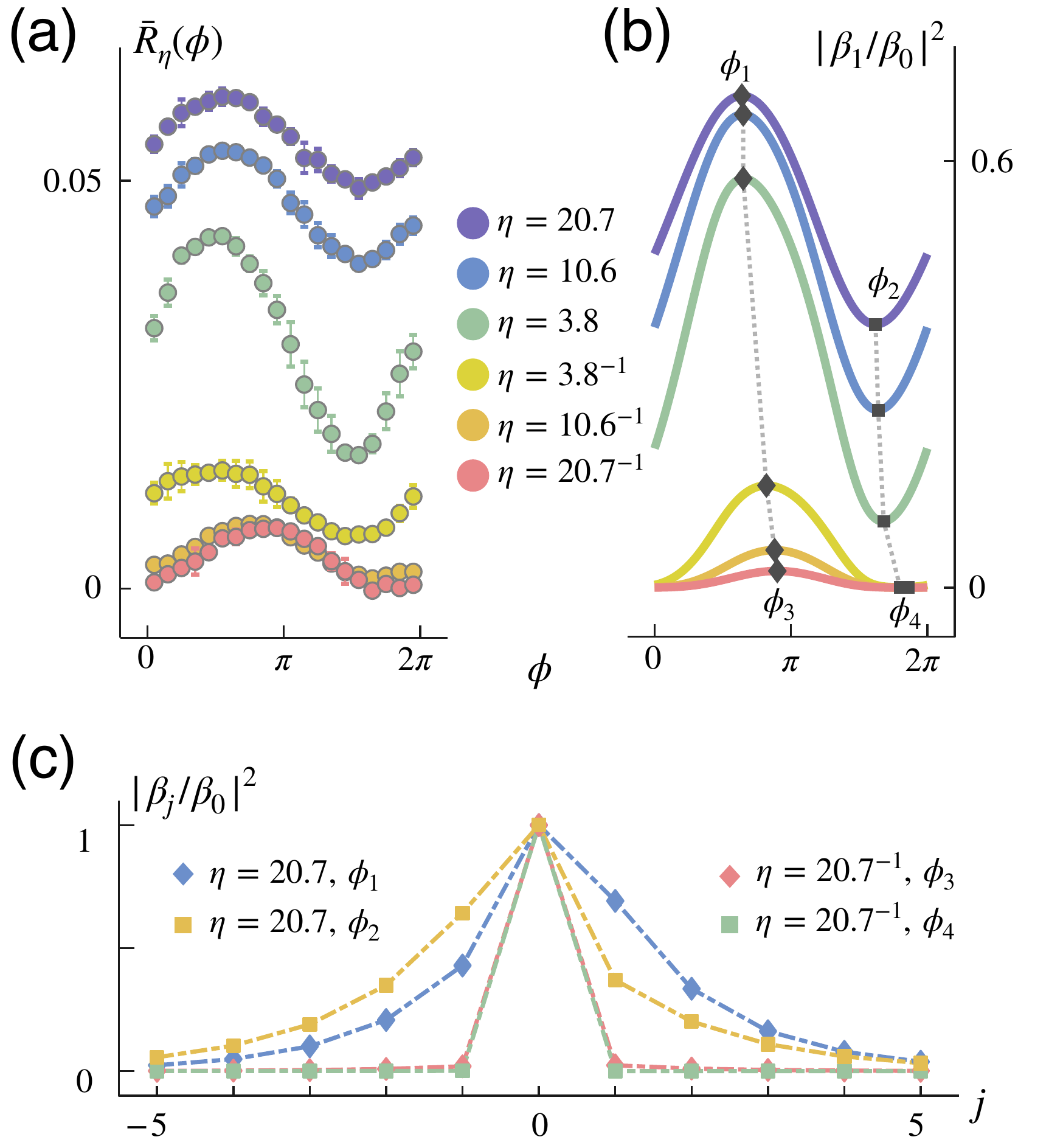}\caption{Response of the dispersive and flat bands with different gauge fields.
(a) The averaged reflectivity $\bar{R}$ versus $\phi$ with different
$\eta$. (b) The normalized probability on the site $b_{1}$ versus
$\phi$. The diamonds (squares) indicate where the curves reach their
maxima (minima). $\Delta_{p}(\Delta_{p}^{\prime})=-2\text{ MHz}$
and $\Delta_{c}=233.5\text{ MHz}$. The gray dashed lines are plotted
to guide the extrema of the curves. (c) The probability distribution
of the steady state $\protect\ket{\psi_{s}}$ on the $b$-leg with
$\phi_{1}=0.64\pi$, $\eta=20.7$ (blue diamonds), $\phi_{2}=1.62\pi$,
$\eta=20.7$ (yellow squares), $\phi_{3}=0.9\pi$, $\eta=1/20.7$
(red diamonds), and $\phi_{4}=1.86\pi$, $\eta=1/20.7$ (green squares).
The powers of each plane wave component of the two standing waves
are $P_{1}=67\text{ mW}$, $P_{2}=90\text{ mW}$ for $\eta=3.8$ and
$1/3.8$; $P_{1}=40\text{ mW}$, $P_{2}=153\text{ mW}$ for for $\eta=10.6$
and $1/10.6$; and $P_{1}=29\text{ mW}$, $P_{2}=215\text{ mW}$ for
$\eta=20.7$ and $1/20.7$. Other experimental parameters are the
same as in Fig.$\ $\ref{fig1}. The points are simply connected for
clarity. Error bars are obtained from four independent data sets.}

\label{fig3}
\end{figure}
The reflection spectrum is mostly contributed by the slowly moving
atoms that have Doppler shifts smaller than the lattice bandwidth
\cite{cai2019}. We take the average of reflectivity over the bands,
i.e., $\bar{R}=\int Rd\nu_{p}/\int d\nu_{p}$, to investigate the
localization and its dependence on $\phi$. In Fig.$\ $\ref{fig3}(a),
the flat-band localization is demonstrated by the suppression of $\bar{R}$
when $\eta$ decreases. Furthermore, we notice that the $\phi$-dependence
of $\bar{R}$ changes with $\eta$ (see the $\phi$ calibration in
\cite{supp}). The sinusoidal curve of averaged reflectivity $\bar{R}_{\eta}(\phi)$
is shifted from top to bottom in Fig.$\ $\ref{fig3}(a).

The $\phi$-dependent shift shows the distinct responses to the gauge
fields of the two bands. We consider two ideal cases to explain the
physics. If the excitation is completely prepared in the dispersive
band, the transport dynamics is determined by the flux-dependent unidirectional
chiral edge current \cite{atala2014,mancini2015,stuhl2015,livi2016,anisimovas2016,an2017,cai2019,dutt2020}
of the dispersive band. The unidirectional chiral current breaks the
symmetry between the transition from $b_{0}$ to $b_{1}$ and the
one from $b_{0}$ to $b_{-1}$. When the magnetic flux $\phi\in(0,\pi)$,
the chiral current enhances the probability in the site $b_{1}$,
and hence the reflectivity increases (vice versa). On the other hand,
flat-band response to the gauge field can be understood by the CLSs
$\ket{F_{j}}$ \citep{flach2014detangling}. When $\phi=2n\pi$ ($n$
is an integer), $\ket{F_{j}}\propto(\eta a_{j}^{\dagger}-b_{j}^{\dagger})\ket G$
is localized within the $j$th unit cell. Therefore, only $\ket{F_{0}}$
is excited when we probe the site $b_{0}$, leading to the maximum
localization. Otherwise, $\ket{F_{j}}\propto(\eta a_{j+1}^{\dagger}-e^{i\phi/2}b_{j+1}^{\dagger}+\eta e^{i\phi/2}a_{j}^{\dagger}-b_{j}^{\dagger})\ket G$
is localized within two unit cells. Probing site $b_{0}$ leads to
a coherent superposition between $\ket{F_{0}}$ and $\ket{F_{-1}}$,
and hence results in a finite overlap with $b_{\pm1}$, which is similar
to the ``breathing motions'' in Refs. \cite{mukherjee2018,kremer2020}.

\begin{figure}
\includegraphics[width=0.8\columnwidth]{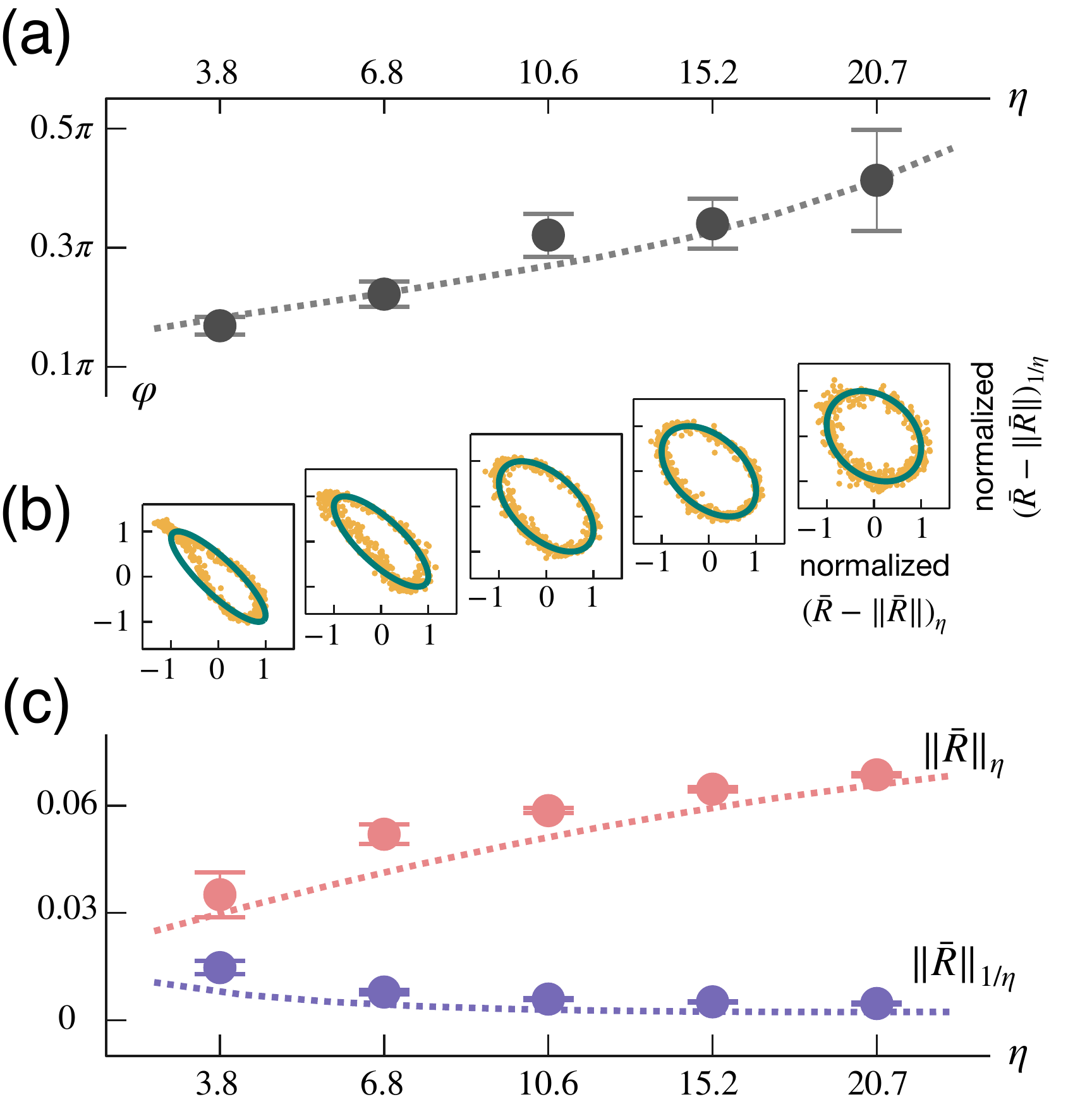}

\caption{Gauge field dependence of the reflectivities from the dispersive band
and flat band. (a) The phase difference $\varphi$ between the $\phi$-dependences
of $\bar{R}_{\eta}$ and $\bar{R}_{1/\eta}$. (b) The Lissajous curves
composed of $\bar{R}_{\eta}(\phi)$ and $\bar{R}_{1/\eta}(-\phi)$.
The phase difference between $\bar{R}_{\eta}$ and $\bar{R}_{1/\eta}$
increases and approaches $\pi/2$ when $\eta$ increases. The yellow
dots are experimental data and the green lines are fitted elliptic
curves. The powers of each plane wave component of the two standing
waves are $P_{1}=50\text{ mW}$, $P_{2}=124\text{ mW}$ for $\eta=6.8$
and $1/6.8$; $P_{1}=33\text{ mW}$, $P_{2}=183\text{ mW}$ for $\eta=15.2$
and $1/15.2$. Other experimental parameters are the same as in Fig.$\ $\ref{fig3}.
Each plot contains 400 data points. (c) The mean averaged reflectivity
$\Vert\bar{R}\Vert_{\eta}$ versus $\eta$. $\Vert\bar{R}\Vert_{\eta}$
increases with $\eta$ monotonically. The dotted lines are the simulated
results.}
\label{fig4}
\end{figure}
The two responses can be further investigated by the steady state
of the collective excited states of atoms, where the wave function
$\ket{\psi_{s}}\approx[1+\sum_{j}(\alpha_{j}a_{j}^{\dagger}+\beta_{j}b_{j}^{\dagger})]\ket G$
$(\alpha_{j},\beta_{j}\ll1)$ in the weak excitation approximation.
In the steady state, we obtain the probability amplitude $\beta_{j}=\sqrt{N}\Omega_{p}\bra Gb_{j}(\Delta_{p}+i\hat{\gamma}-H_{s})^{-1}b_{0}^{\dagger}\ket G$
\citep{ozawa2014anomalous}, where $\hat{\gamma}=\sum_{j}(\gamma_{a}a_{j}^{\dagger}a_{j}+\gamma_{b}b_{j}^{\dagger}b_{j})$
and $\gamma_{a}$ ($\gamma_{b}$) is the decoherence rate between
the excited state $\ket a$ ($\ket b$) and $\ket g$. Especially,
we plot $\left|\beta_{1}/\beta_{0}\right|^{2}$, i.e., the normalized
probability on the site $b_{1}$ in Fig.$\ $\ref{fig3}(b). Since
the reflectivity is approximately proportional to $\left|\beta_{1}/\beta_{0}\right|^{2}$
\citep{Wang2015}, both features of the $\bar{R}$ in Fig.$\ $\ref{fig3}(a)
are demonstrated, including the magnitude suppression and the $\phi$-dependence.
We also plot four typical probability amplitude distributions along
the $b$-leg in Fig.$\ $\ref{fig3}(c) with $\eta\gg1$ and $\eta\ll1$.
For $\eta=20.7$, the distributions of $\vert\beta_{j}\vert^{2}$
with $\phi_{1}=0.64\pi$ and $\phi_{2}=1.62\pi$ are almost symmetric
to each other with respect to the $0$th site, implying flux-dependent
unidirectional chiral edge current of the dispersive band. However,
the distributions with $\phi_{3}=0.9\pi$ and $\phi_{4}=1.86\pi$
are symmetrically localized at $b_{0}$ with different localization
lengths. $\left|\beta_{1}/\beta_{0}\right|^{2}$ for $\eta=1/20.7$
is minimized at $\phi_{4}$, which means that the eigenstates are
maximally localized when the synthetic gauge field almost vanishes.
We notice that $\phi_{4}$ is slightly different from $2n\pi$, which
comes from the suppressed but finite contribution from the dispersive
band. 

In Fig.$\ $\ref{fig4}(b), the different $\phi$-dependences of the
dispersive and flat bands are illustrated with Lissajous curves. The
spatial phase difference between the envelopes of the two standing
waves is slowly tuned to cover all values of $\phi$. We obtain the
data sets \{$\bar{R}_{\eta}(\phi),\bar{R}_{1/\eta}(-\phi)$\} and
fit them with ellipses to reconstruct the Lissajous curves. The shape
of the ellipse elucidates the phase differences between the two underlying
functions. For example, a Lissajous curve composed of two parametric
equations with argument $u$, e.g., $x=\sin(u)$ and $y=\sin(-u+\varphi)$,
is a line (circle) when $\varphi=0$ ($\pi/2$). In Fig.$\ $\ref{fig4}(a),
$\varphi$ obtained by fitting the Lissajous curve approaches $\pi/2$
when $\eta$ increases, indicating the different types of the excitations
on the flat and dispersive bands. In Fig.$\ $\ref{fig4}(c), we notice
that $\Vert\bar{R}\Vert_{\eta}$ increases with $\eta$ monotonically,
where $\Vert.\Vert$ indicates the mean value of the averaged reflectivity
$\bar{R}$ over all $\phi$. The numerical simulation agrees with
the data. 

In conclusion, we experimentally realize Creatz ladders with tunable
gauge fields, where the flat band can be selectively excited and the
interplay between the flatband localization and the AB phases was
investigated. We study the flat-band localization in an open system,
where the steady state balanced by pumping, driving, and dissipation
exhibits the dynamics in the corresponding closed system \citep{ozawa2014anomalous}.
We also need to emphasize that our scheme is substantially different
from the incoherently pumped polariton-exciton condenstates \cite{baboux2016,harder2020,masumoto2012,klembt2017,whittaker2018},
where coherence is not accessible between multiple CLSs. It is interesting
to notice that both bands of the Creutz ladder are topologically non-trivial
\cite{kremer2020,kang2020} provided that $\phi\neq2n\pi$ \cite{supp}.
It is a step towards the simulation of the strong correlated quantum
phases, including the fractional Chern insulators \cite{bergholtz2013},
disorder-free many-body localization \cite{kuno2020}, and unusual
ferromagnetism \cite{tasaki1992}. An interaction term between the
sites in momentum space can be introduced by weakly coupling the excited
atomic level to a Rydberg state \cite{li2020} or by $s$-wave interactions
\cite{an2018,xie2020}. With a negative $\Delta_{c}$, the flat band has
the lowest energy and can be used to study the many-body ground states
of ultracold atoms \cite{li2020,supp}.

We acknowledge the support from the National Natural Science Foundation
of China (Grants No. 11934011, No. 11874322, No. 91736209 and No.
U1330203), the National Key Research and Development Program of China
(Grants No. 2019YFA0308100 and No. 2018YFA0307200), and the Fundamental
Research Funds for the Central Universities.

$\ $

$^{*}$These authors contributed equally to this work.

$^{\dagger}$hancai@zju.edu.cn

$^{\dagger\dagger}$junxiang$\_$zhang@zju.edu.cn

\end{document}